\documentstyle[12pt]{article}

\title{\bf The Aleph Cosmological Principle$^1$}

\author{Domingos Soares \\ \\ {Physics Department}\\
{Federal University of Minas Gerais} \\   
{Belo Horizonte, Brazil} } 

\date{September 1, 2004}

\begin{document}
\maketitle

\hfill `I saw the Aleph from every point and angle, 

\hfill and in the Aleph I saw the earth and in the earth the Aleph...' 

\hfill {\it The Aleph}, Jorge Luis Borges, 1945\\

%
% ...because \footnote cannot be used in the title!!!
% Use \footnotemark only to generate the footnote mark and \footnotetext 
% to generate the text. See old manual, page 156
%                                                  by dsoares, 11jun04
%
\addtocounter{footnote}{1}
\footnotetext{For more short comments on modern cosmology check 
at the following address:
http://lilith.fisica.ufmg.br/dsoares/notices-e.htm}

\begin{abstract}
A general cosmological principle --- Aleph --- is proposed as a 
substitute to the Anthropic principle. Furthermore, the universe, 
conceived as a world ensemble, is 
characterized by many (possibly infinite) X-Life world 
principles. The only known X-Life world principle recovers much of the 
Anthropic conjecture. The inescapable final conclusion is the formulation 
of the Strong Copernicus principle.
\end{abstract}

\bigskip\bigskip\bigskip

\section{Introduction}
The Anthropic cosmological principle (Carter 1974, Barrow \& Tipler 1986) 
has been criticized, and eventually rejected as inadequate by some authors, 
for being heavily {\it inspired} on unproved cosmological models, namely, 
those known as Hot Big Bang models (Soares 2004a).

Carter presented his Anthropic principle in two versions, weak and strong, 
whilst Barrow \& Tipler described other versions. The great novelty lies 
in the weak version. The discussion that is done here focuses, 
therefore, upon the weak version of the principle.

In fact --- and it is worth-stressing ---, the different versions of the 
Anthropic Principle are not different versions of the same principle 
but rather are {\it independent principles by themselves,} which is 
totally opposed to the view expressed mainly by Barrow \& Tipler. Such 
a thesis is further elaborated elsewhere (Soares 2004b). 

Towards a broader and unprejudiced view, one can depart from the idea of 
a universe as a  {\it world ensemble} (e.g., Carter 1974), except that 
in a different perspective from what is usually found in the literature,
namely, that of a {\it multiverse} (see details in Stoeger et al. 2004 
and references therein). Let each world vector --- i.e., 
each element in the ensemble ---, in fact, 
belong to the {\it same} universe, not being 
a {\it universe} by itself, with its own cosmology, as assumed in the 
usual world-ensemble approach. That 
is to say, the total mass-energy content of the universe is given by 
adding up the mass-energy content of each world vector. Furthermore, each 
world vector is assumed as potentially suitable for the existence --- or 
development --- of life. In other words, the overall conditions in that --- 
and all --- world element are such that live organisms are bound to emerge. 
It is thus characterized by a {\it X-Life world principle}, which simply 
states that world is as it is because of restrictions imposed by X-Life being 
the way it is. These are essentially the same words in Carter's formulation 
of his Anthropic principle. Here they are used in the context of a much 
broader cosmological view as it will be apparent below.

The Anthropic principle has been used in many ways since its proposition. 
A sort of strange {\it devotion} sometimes characterizes those dealing 
with the principle. This resulted into an exaggerated {\it bending 
of the bow} towards one direction. The situation, comprehensibly, 
led Soares (2004a) to use irony on the whole issue, in an attempt to bend 
the bow to the opposite direction, eventually reaching a state of 
reasonable equilibrium. That is also the spirit pervading the present 
essay, except that now with a grave approach.

\section{The E-Life world principle}

Ours --- the only presently {\it applicable} X-Life world principle ---, 
conveniently, could be termed {\it E-Life world principle}, where 
"E" stands obviously for "Earth". Since the DNA-molecule 
is the unifying feature of terrestrial life, the principle 
is thus  stated as constraints derived from DNA-based life upon the world 
vector properties. Very much so, it is the present general idea pervading the 
Anthropic principle. 

Except for the cosmological implications, much of the 
conclusions derived from the Anthropic principle (Barrow \& Tipler 1986)  
surely still holds. For example, the prediction by Fred Hoyle concerning 
the 7.7 MeV excited state of $^{12}$C, which was necessary in order to 
increase the probability of the reaction between helium and beryllium 
to produce carbon, might be considered as a genuine E-Life world principle 
prediction. In 1952, from the evident abundance of carbon --- 
namely, E-Life ---, Hoyle predicted the existence of a resonance 
of $^{12}$C, in nuclear reactions, at around 7.7 MeV, and almost 
immediately, in 1953, D.N.F. Dunbar, R.F. Pixley, W.A. Wenzel \& 
W. Whaling (1953), at Kellogg Radiation Laboratory, Caltech, 
discovered a state with the correct properties, at 7.68$\pm$0.03 MeV 
excitation energy. E-Life would not exist without the 7.7-MeV excited 
state of $^{12}$C. Such a prediction is, of course, often mentioned 
in classic Anthropic discussions (see Barrow \& Tipler, p. 252).

The reason why general cosmological implications are not valid is that a given 
cosmology must be applied to the whole world ensemble and not to a sole 
element of it. {\it Aleph} is the applicable principle here (see below). 
{\it Cosmological predictions are always biased when based in a X-Life world 
principle.}

Intelligent life is always an issue whenever one speaks of {\it life}. 
Intelligence, another variable in the general cosmological equation, is 
not considered in the present discussion. Irrespective of 
its prevalence, communications between world-ensemble elements, e.g., 
between a particular X-based organism and a DNA-based one, may or may 
not be possible. 
In any case, whether or not two elements of the world ensemble 
are or may be connected in one or other way --- communication being one 
of them --- is entirely irrelevant here. 

\section{A hypothetical X-Life world principle}
Sagan \& Salpeter (1976) discuss many aspects of a possible Jovian 
biology, an investigation motivated mainly by the fact that contemporary 
Jovian atmosphere has many similarities to the primitive terrestrial 
atmosphere. They hypothesized the characteristics of Jovian live organisms --- 
in the form of sinkers and floaters, understandable in a gaseous 
environment --- departing from chemical composition, temperature, 
density, pressure and other known features of the planet atmosphere. 
Fundamental for the origin of life in Jupiter is the time-scale taken 
by synthesized complex molecules to move towards large depths, as a 
result of convective streaming. The time-scale should be short enough 
to avoid reaching pyrolytic depths, which would severe restrict 
the possibility of biological evolution. 

Now, take the Sagan-Salpeter problem in the reverse order. Assume 
sinkers and floaters {\it are} abundant in the Jovian atmosphere. 
One may then formulate the {\it J-Life} world principle --- 
"J" for "Ju\-piter". With such a life principle, properties of the Jovian 
atmosphere might be obtained in the same way E-Life --- Anthropic --- 
predictions are made.

Incidentally, Jovian live balloons would be, in principle, totally 
disconnected from DNA-based terrestrial life. In other words, J-Life 
world and E-Life world would be disconnected from each other. 

I am assuming here that Jovian organisms are not DNA-based, which may 
not be true. But that does not invalidate the example.

\section{Aleph and Copernicus principles}
The Aleph Cosmological Principle is the underlying principle to the world 
ensemble, i.e., to the universe. Predictions about the formation, evolution 
and structure of the universe --- cosmology --- are related to the Aleph 
principle. It is much more general in scope than each X-Life world principle.

Strictly speaking, X-Life world principles do not need {\it intelligent}  
life to hold. In particular, E-Life world principle would still hold 
even {\it in the absence of mankind}, of human beings. This is the essence 
of what could be termed {\it the Strong Copernicus principle}. As long 
as any sort of DNA-based life do exist, the E-Life principle would still 
be there. Useless, due to the absence of intelligent life. But still there. 
The Strong Copernicus principle does not require the existence of human 
beings. Man is not central in the universe, it may even not exist. Putting 
it in another form, amongst all DNA-based forms of life, man is not special. 
Consciousness makes mankind {\it different} --- not special --- from other 
E-life organisms, in the sense that mankind is dotted with moral 
and ethical values. These are fundamental aspects of human life but do not 
change the biological status of human life. In conclusion, 
the Strong Copernicus principle is an imperative {\it scientific}  
principle. 

The philosophical implications herein are innumerable and will not be 
treated here. 

\section{Conclusion}
At this point in time there is only one X-Life principle known. 
Conceivably, the Aleph cosmological principle is a matter of speculation. 
Conceptually, however, it leads to a broader scenario for the knowledge of 
the universe we live.

Soares (2001) suggests that the arrow of time is given by the prominence of 
life, that is, the universe evolves towards life. In an eternal universe, 
that would lead to the startling conclusion that the {\it universe itself 
is alive!} The Aleph principle is, then, at a certain point, vindicated.

The Aleph cosmological principle is the Aleph-Life principle. It is of course 
prompted to speculation what is the nature of Aleph-Life, in a way or other, 
the live universe. Highly speculative matter, on the other hand, 
scientifically unavoidable.

\section{References}
\begin{description}
\item Barrow, J.D. \& Tipler, F.J. 1986, The Anthropic cosmological
principle (Oxford University Press, Oxford)
\item Carter, B. 1974, Large number coincidences and the Anthropic principle 
in cosmology, in Confrontation of cosmological theories with observational 
data, IAU Symposium No. 63, Krakow, Poland, September 10-12, 1973.
ed. M.S. Longair, D. Reidel Publishing,  Dordrecht, p. 291-298
\item Dunbar, D.N.F., Pixley, R.E., Wenzel, W.A. \& Whaling, W. 1953, 
The 7.68-Mev State in C$^{12}$. Phys. Rev., 92, 649-650
\item Sagan, C. \& Salpeter, E.E. 1976, Particles, environments, and 
possible ecologies in the Jovian atmosphere. Astrophys. J. Supp., 32, 737-755.
\item Soares, D.S.L. 2001, Time is life, 
http://arXiv.org/abs/astro-ph/0108180
\item Soares, D.S.L. 2004a, The Anthropic Fake Principle, \\
http://www.fisica.ufmg.br/\char125 dsoares/antr/fake.htm
\item Soares, D.S.L. 2004b, In preparation
\item Stoeger, W.R., Ellis, G.F.R. \& Kirchner, U. 2004, Multiverses and Cosmology: 
Philosophical Issues, http://arXiv.org/abs/astro-ph/0407329 
\end{description}

\end{document}